*Title:*

# Threshold Reaction Rates and Energy Spectra of Neutrons in the 0.8-1.6 GeV Proton-IrradiatedW, Na Targets

*Author(s):*

Yury E. TITARENKO, Vyacheslav F. BATYAEV, Evgeny I. KARPIKHIN, Valery M. ZHIVUN, Svetlana V. KVASOVA, Ruslan D. MULAMBETOV, Dmitry V. FISCHENKO, Aleksander B. KOLDOBSKY, Yury V. TREBUKHOVSKY, Vladimir A. KOROLEV, Gennady N. SMIRNOV, Andrey M. VOLOSHENKO, Vladimir Yu. BELOV, Nikolay I. KACHALIN, Stepan G. MASHNIK, Richard E. PRAEL, Arnold J. SIERK, Hideshi YASUDA

# Threshold Reaction Rates and Energy Spectra of Neutrons in the 0.8-1.6 GeV Proton-Irradiated W, Na Targets

Yury E. TITARENKO[1,*], Vyacheslav F. BATYAEV[1], Evgeny I. KARPIKHIN[1], Valery M. ZHIVUN[1], Svetlana V. KVASOVA[1], Ruslan D. MULAMBETOV[1], Dmitry V. FISCHENKO[1], Aleksander B. KOLDOBSKY[1], Yury V. TREBUKHOVSKY[1], Vladimir A. KOROLEV[1], Gennady N. SMIRNOV[1], Andrey M. VOLOSHENKO[2], Vladimir Yu. BELOV[3], Nikolay I. KACHALIN[3], Stepan G. MASHNIK[4], Richard E. PRAEL[4], Arnold J. SIERK[4], Hideshi YASUDA[5]

[1] *Institute for Theoretical and Experimental Physics, B.Cheremushkinskaya 25, 117259 Moscow, Russia*

[2] *Keldysh Institute of Applied Mathematics, Miusskaya Sq. 4, 125047 Moscow, Russia*

[3] *Russian Federal Nuclear Center – VNIIEF, 607190, Sarov (Arzamas-16), Nizhny Novgorod region, Russia*

[4] *Los Alamos National Laboratory, Los Alamos, NM 87545, USA*

[5] *Japan Atomic Energy Research Institute, Tokai-mura, Ibaraki 319-1195, Japan*

Considering the prospects of using the W-Na target assemblies in ADS facilities, the experiments were made to study the nuclear-physics characteristics of W and Na, and the composite structures thereof in their interactions with 0.8-GeV and 1.6-GeV protons. The neutron and proton-induced reaction rates were measured inside, and on the surface of, a cylinder-shaped heterogeneous W-Na assembly together with the double-differential spectra of secondary neutrons emitted from different-depth W and Na discs. The measurement results were simulated by the LAHET, CEM2k, and KASKAD-S codes in terms of the latest versions of nuclear databases.

## I. Introduction

Tungsten is regarded as belonging to the most promising class of target materials in the present-day conceptual designs of ADS facilities. Tungsten is sufficiently easy to handle, is of a high density, shows the desired set of nuclear-physics characteristics, and does not exhibit chemical reactivity and biological toxicity (contrary to Pb and Hg).

At the same time, the actual target designs include not only the neutron sources made of heavy materials, but also heat exchangers with coolants. Na is often proposed to be a coolant, considering that the Na-based technologies have been studied quite properly. Besides, Na can also be used in the heterogeneous target designs to level the energy deposit within the target volume.

In view of the above, accumulation and estimation of the microscopic and group nuclear constants for the said materials and for the assemblies thereof have become urgent. Bearing in mind that the target structures are very complicated and can be subject to ample alterations, the computational methods based on the latest simulation codes and nuclear databases can only be used for that purpose. On the other hand, as mentioned repeatedly in the relevant publications (see[1)] , for instance), the present-day codes are often devoid of the predictive power required by the actual ADS facility designs.

A necessity has arisen, therefore, for the simulation codes to be tested by the results of experimenting with the prototypes of the target parts and units. It should not be forgotten either that the said experimental results can, and certainly must, be used as independent nuclear constants. The present work describes two groups of the experiments, namely,

- studying the threshold reaction rates inside, and on the surface of, a 0.8 GeV proton-irradiated lamellate heterogeneous W-Na target;
- measuring the double-differential neutron spectra from 0.8 GeV and 1.6 GeV proton-irradiated W and Na targets.

The experimental results were simulated by the LAHET code (in both cases), by the KASKAD-S code (the threshold reaction rates), and by the CEM2k code (the neutron spectra).

## II. Experiment

The experiments were made with the ITEP U-10 proton synchrotron beams of $7.2{\cdot}10^{10}$ p/pulse (the reaction rates) and $\sim 1{\cdot}10^5$ p/pulse (the secondary neutron spectra) intensities.

The general experimental design of measuring the reaction rates has been described in[2)] . The measurements were made by the gamma-spectrometry method using 142 experimental samples of thirteen single-isotope and high isotopic enriched materials ($^{12}$C, $^{19}$F, $^{27}$Al, $^{59}$Co, $^{63}$Cu, $^{65}$Cu, $^{64}$Zn, $^{93}$Nb, $^{115}$In, $^{169}$Tm, $^{181}$Ta, $^{197}$Au, and $^{209}$Bi). The 0.03-2.1 g/cm$^2$ disc-shaped samples were placed at 84 points inside, and on the surface of, a heterogeneous W-Na assembly. The continuous irradiation time was 10 hours, with a $6.5{\cdot}10^{14}$ integral proton fluence. After the irradiation, the samples were gamma-spectrometered using the GC2518 Ge detector of a 1.8 keV resolution in the 1332 keV gamma-line. The resultant gamma-spectra were processed by the GENIE2000 code. The secondary nuclide were identified by the ITEP-developed SIGMA code using the PCNUDAT database.

The double-differential spectra of secondary neutrons were measured by the techniques[3)] using a TOF spectrometer. The neutrons were recorded with the BICRON 5MAB1F D12.7*L15.2cm detectors based on BC501 liquid scintillator. The projectile proton energies were 0.8 GeV and 1.6 GeV; the neutrons were detected at 30$^o$, 60$^o$, 90$^o$, 120$^o$, and 150$^o$. The gamma-background was discriminated by the pulse shape. The absolute neutron detection efficiency was calculated by the SCINFUL[4)] and CECIL[5)] codes.

[*] Corresponding author, Tel. +7-095-123-6383, Fax. +7- 095-127 -0543, E-mail: Yury.Titarenko@itep.ru

## III. Computational simulation and comparison with experimental data

The experimental data were simulated by the LSC code system[6] using the LAHET code (in both cases), the KASKAD-S code[7] (threshold reaction rates), and the CEM2k code[8] (neutron spectra). The simulation techniques are described in detail in[2] .

The results of comparing between the experimental and simulated data are presented in Figs. 1–3 (reaction rates) and 4–5 (secondary neutron spectra). The mean squared deviation factor $<F>$ was used as a criterion for agreement between the calculated and experimental reaction rates.

Beside the comparison between the experimental and simulated results of the present work, Fig. 5 shows the results of work[9] , which is the only work that admits a straightforward comparison with our data.

## IV. Conclusions

The experimental data and the results of testing the simulation codes have permitted the following conclusions:

- many of the measured reactions cannot be reproduced by simulation in terms of the present-day cross section libraries (MENDL2 and MENDL2P) because of their restricted energy range[10] ($E_n$<100MeV, $E_p$<200MeV);

- the results obtained indicate that the cumulating effect in the high-energy neutron reactions has to be allowed for;

- the studied reaction rates inside, and on the surface of, a heterogeneous W-Na target indicate that the codes based on Monte-Carlo techniques (e.g., LAHET) are much more promising compared with the codes based on the discrete-ordinate algorithm (e.g., KASKAD-S);

- theoretical models for separating the secondary product yields between the isomeric states have urgently to be constructed with a view to updating the simulation codes; the absence of such a model prevents from a reliable simulation-to-experimental comparison for a number of reactions ($^{115}$In(n,p)$^{115g}$Cd, for instance);

- the quality of the calculated reproduction of the double-differential spectra of secondary neutrons is much higher in the case of W compared with Na, the fact that can qualitatively be explained by the drawbacks of the present-day theoretical models for reactions with small-mass nuclei at intermediate and high energies;

- additional measurements of the space-energy characteristics of secondary neutrons seems to be urgent because the available data are scanty and preclude any final conclusion concerning the reasons for the discrepancies in the experimental results of the given type; moreover, a nuclear database that could have been safely used in practice is impossible to compile using the present-day experimental data.

### Acknowledgment

The authors are indebted to Dr. S. Meigo (JAERI) for his assistance in calculating the detection efficiency, to Dr. V. L. Romodanov (MEPI, Moscow) for his assistance in calibrating the detectors, to Drs. V. N. Kostromin and I. A. Vorontsov (ITEP, Moscow) for their assistance in carrying out the experiments, and to Dr. F. E. Chukreev (KIAE, Moscow) for discussion of the results.

The work has been carried out under the ISTC Project #1145 supported by JAERI (Japan). In part, the work has been supported by the U.S. Department of Energy.

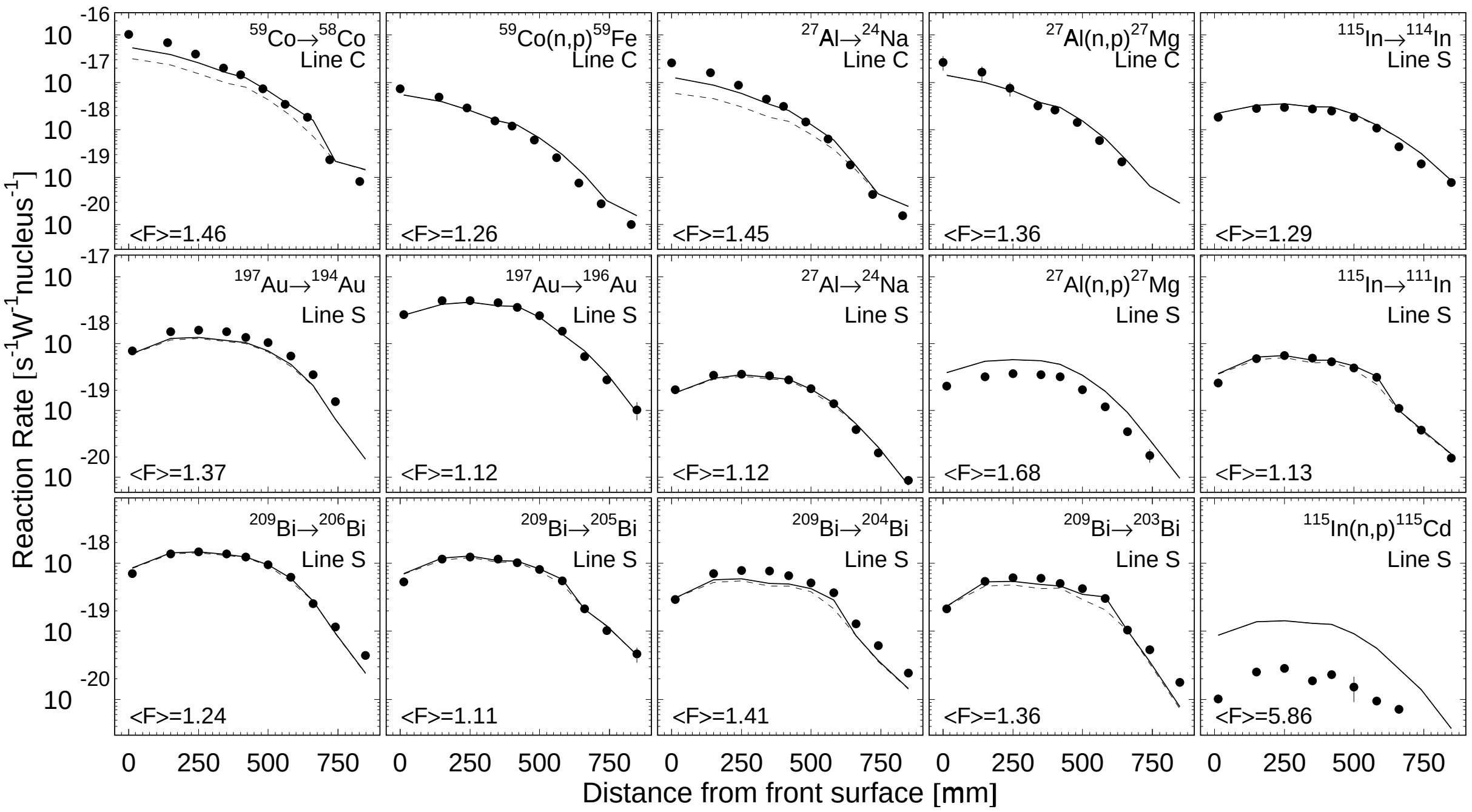

**Fig. 1** The experimental and the LCS calculated reaction rates. The dashed line is the neutron contribution to the reaction rates. The mean squared deviation factor <F> is also shown. The presented reaction rates have been normalized to the proton beam power.

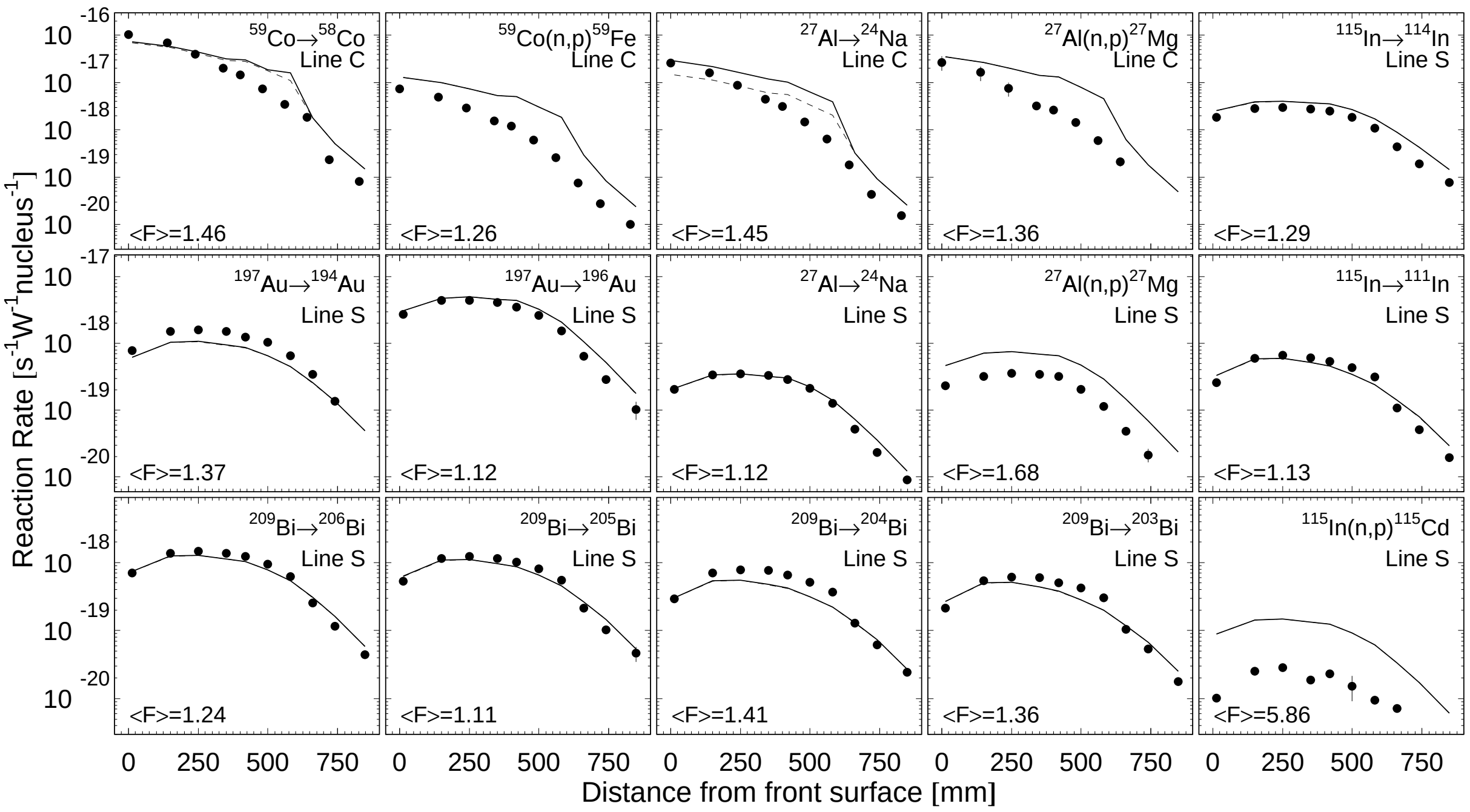

**Fig. 2** The experimental and the KASKAD-S calculated reaction rates. The dashed line is the neutron contribution to the reaction rates. The mean squared deviation factor <F> is also shown. The presented reaction rates have been normalized to the proton beam power.

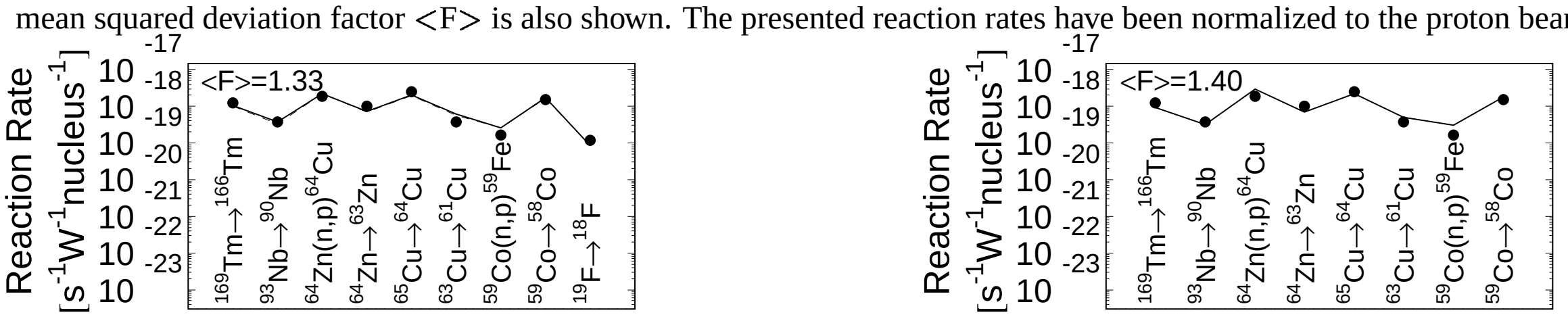

**Fig. 3** The experimental and the LCS (left figure) and KASKAD-S (right figure) calculated reaction rates at the surface of second W-disk. The dashed line is the neutron contribution to the reaction rates. The mean squared deviation factor <F> is also shown. The presented reaction rates have been normalized to the proton beam power.

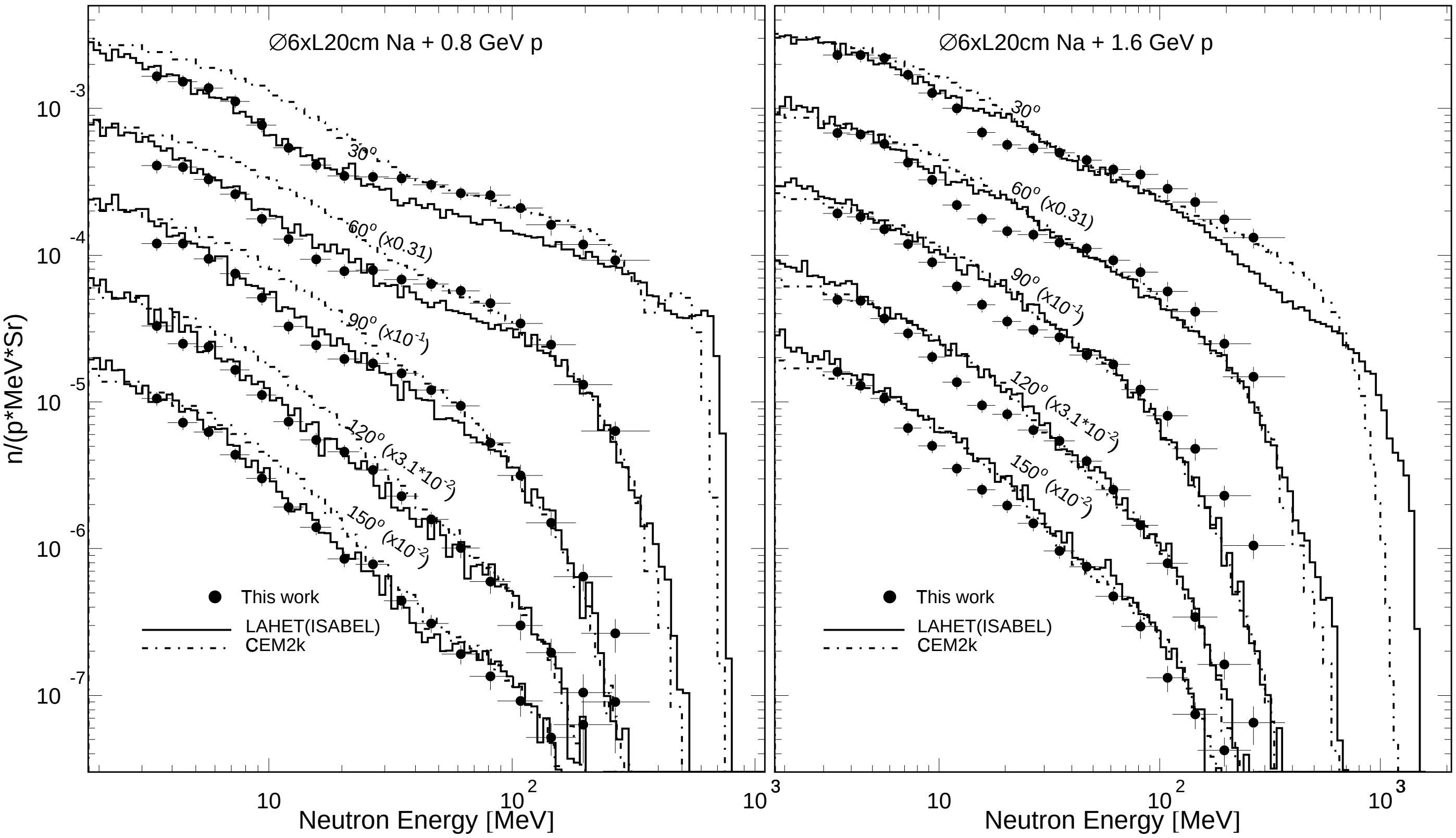

**Fig. 4** The double differential neutron spectra measured in the present work from D6xL20 cm Na irradiated by the 0.8 (the left-hand panel) and 1.6 (the right-hand panel) GeV protons, together with the LAHET[6)] and CEM2k[8)] code calculation results.

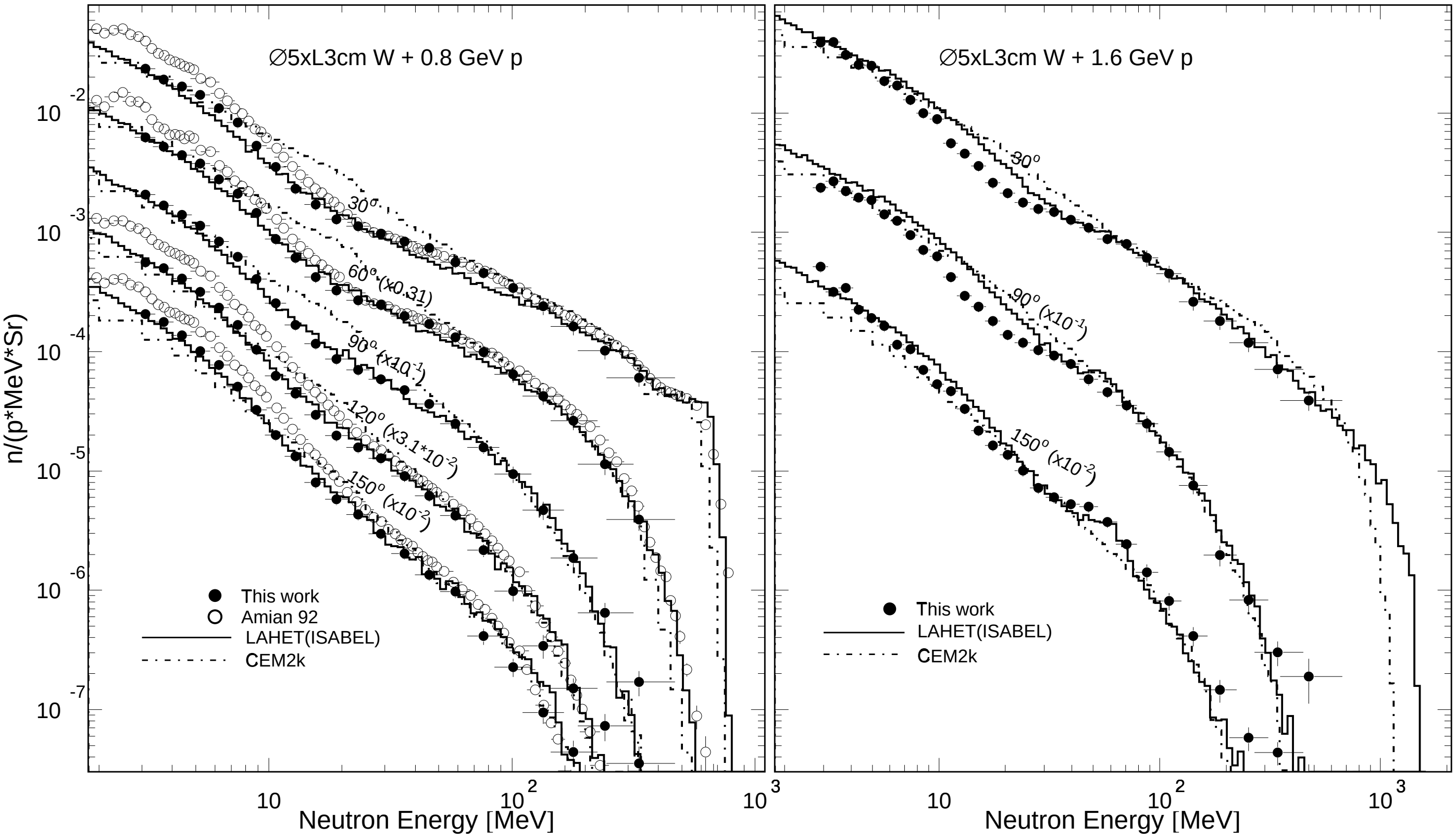

**Fig. 5** The double differential neutron spectra from D5xL2.935 cm W irradiated by the 0.8 GeV (the left-hand panel) and 1.6 GeV (the right-hand panel) protons, as measured in the present work and in[9)] for 0.8 GeV protons, together with the LAHET[6)] and CEM2k[8)] code calculation results.